# Tapered semiconductor amplifiers for optical frequency combs in the near infrared

Flavio C. Cruz[1,2], Matthew C. Stowe[1], and Jun Ye[1]

[1] JILA, National Institute of Standards and Technology and University of Colorado
Department of Physics, University of Colorado, Boulder, CO 80309-0440
[2] Instituto de Física Gleb Wataghin,Universidade Estadual de Campinas, CP.6165, Campinas, SP, 13083-970, Brazil

A tapered semiconductor amplifier is injection seeded by a femtosecond optical frequency comb at 780 nm from a mode-locked Ti:sapphire laser. Energy gains over 17 dB (12 dB) are obtained for 1 mW (20 mW) of average input power when the input pulses are stretched into the picosecond range. A spectral window of supercontinuum light generated in a photonic fiber has also been amplified. Interferometric measurements show sub-hertz linewidths for a heterodyne beat between the input and amplified comb components, yielding no detectable phase-noise degradation under amplification. These amplifiers can be used to boost the infrared power in f-to-2f interferometers used to determine the carrier-to-envelope offset frequency, with clear advantages for stabilization of octave-spanning femtosecond lasers and other supercontinuum light sources with very limited power in the infrared.

Tapered semiconductor amplifiers (TSA) are known to provide large gains ($\geq$ 20 dB) for near infrared continuous-wave (cw) lasers, converting a few milliwatts of cw radiation to more than 1 W, without degradation in linewidth[1]. These amplifiers have been widely used in atomic spectroscopy[1,2,3], laser cooling and trapping[4], and for increasing the efficiency of nonlinear frequency conversion of cw lasers into the blue and UV regions[3,4,5]. Amplification of picosecond pulses with semiconductor amplifiers has also been the subject of numerous investigations[6,7], particularly in connection with optical communications. Semiconductor amplifiers enjoy inherent advantages of compactness, optical integration, and reduced cost. In this Letter we show that TSA can be used to amplify short pulse trains from a femtosecond frequency comb[8], while preserving the optical phase coherence. To overcome problems related to gain saturation and finite carrier recombination time in semiconductors (~ 100s of picoseconds), the pulses need to be pre-stretched (chirped)[9]. These features suggest their use for precise optical time transfer[10,11,12] over fiber networks. In addition, we have amplified a spectral window of supercontinuum light generated in a photonic fiber, with gains approaching the unsaturated cw gain at low input powers. TSA can thus aid f-to-2f interferometers used to measure and stabilize the carrier-to-envelope offset frequency of mode-locked lasers[8]. This can be of particular advantage for phase stabilization of octave-spanning femtosecond lasers[13,14] or other supercontinuum near IR laser systems that suffer from limited power in the long-wavelength portion of their spectrum.

With its temperature stabilized to 22 $^0$C, the tapered semiconductor amplifier[15] delivers an output power of 500 mW at 780 nm for an injected current of 1.5 A. The amplifier has a maximum injection current of 3.0 A, with input and output apertures of 3 and 190 μm, respectively, and a length of 2750 μm. The small input aperture requires careful mode matching, although the circular input mode is not shaped to the elliptical mode of the TSA. The amplifier performance is tested under both cw and short pulse injection. A single-frequency diode laser at 780 nm is used as a cw light source. Short pulses are from a mode-locked Ti:sapphire laser, which emits 25 fs pulses at 100 MHz repetition rate, in a transform-limited bandwidth (BW) of 35 nm (FWHM) centered at 780 nm. These pulses have been chirped up to 400 ps by passing through different lengths of optical fibers. Figure 1 shows normalized spectra for the Ti:sapphire laser before and after passing through a 21-m long fiber, which chirps the pulses to 150 ps and introduces some spectral broadening (BW=44 nm). Also shown is the spectrum of the corresponding amplified pulses, showing a gain BW of 14 nm. Spectral components outside the amplifier bandwidth are strongly attenuated. In our measurements of gain, pulse shape, spectra, and phase coherence, we also use a 3-nm bandpass filter at 780 nm (Fig. 1) to limit the laser bandwidth within the amplifier bandwidth. Input and output pulse shapes are measured by using a fast photodetector (15-ps response time) and a 12-GHz sampling oscilloscope. Amplification of chirped pulses in our TSA differs from chirped pulse amplification (CPA) schemes because the amplifier is saturated and its gain bandwidth is smaller than the bandwidth of the pulse to be amplified. CPA has been recently demonstrated with TSA by chirping picosecond pulses to several nanoseconds[9].

Figure 2 summarizes the results for amplification of cw and short pulses as a function of the input power (pulse energy) and pulse duration (chirp). Using the cw laser, or spectrally filtered light from the mode-locked laser, transparency is achieved for a bias current of 670 mA. For cw injection and a current of 2.0 A, an unsaturated (small signal) gain of 20 dB is measured, which reduces to 14 dB for an input power of 20 mW. From the cw gain curve, the saturation power $P_{sat}$ is estimated to be 48 mW, in good agreement with the TSA specifications[15]. 18 mW (42 mW) of input cw power is amplified to 500 mW at the output under a bias current of 2.0 A (1.7 A). For short pulse injection, we measure the average input and output powers. When the laser bandwidth exceeds the amplifier gain bandwidth, the input power is corrected accordingly. In all gain values shown in Fig. 2, the amplified spontaneous emission power (19±1 mW at 2.0 A, without input light) has been subtracted from the output power, which may lead to some underestimation of gain at higher input powers.

Semiconductor amplifiers for picosecond pulses have been the subject of numerous investigations[6,7,16,17,18]. The carrier dynamics is complex, with several processes occurring both at ultrafast (fs) and slow time scales ($\leq$ ns). The main limitation for amplifying short pulses arises from the carrier recombination time $\tau_c$ of a few hundred picoseconds[6]. A strong and fast pulse can quickly reduce the carrier density and deplete the gain, which then takes hundreds of picoseconds to recover. In this way, the trailing edge of a pulse can experience a substantially reduced gain. This gain variation also causes index of refraction changes responsible for self phase modulation (SPM)[6]. It distorts both the pulse shape and spectrum, which in general become asymmetrical[6] (Figs. 1 and 2). This distortion depends on the amplifier gain, initial chirp, pulse duration and input energy ($E_{in}$) compared to the saturation energy ($E_{sat}$)[7]. If each pulse considerably reduces the gain, the recombination time will also limit the repetition rate of amplified pulses to a few GigaHertz. For femtosecond pulses, another important time scale on the order of several hundred fs associated with non-equilibrium ultrafast carrier heating dynamics is responsible for lower values of $E_{sat}$[16], making the gain at increased input power to be much lower than for longer pulses.

Figure 2 shows the gain for 25-fs transform-limited (TL) pulses and for pulses that have been chirped by fibers to 150 and 320 ps. The TSA BW (14 nm), smaller in comparison to the pulse BW, converts these input pulses to 62 fs, 47 ps, and 101 ps, respectively. It is expected that the gain for pulses with duration $\tau_p$ approaches the cw gain if $\tau_p \geq \tau_c$ and $E_{in} \ll E_{sat}$ [7]. We applied the simplified model of Ref. 7, extended to include spontaneous emission [17] and chirped input pulses, in an attempt to account for the results shown in Fig. 2. The asymmetric pulse shapes (inset of Fig.2) are reproduced, with their durations shorter than the input due to spectral filtering of the amplifier gain BW. Although individual curves in Fig. 2 can be reproduced, assuming reasonable values of $E_{sat}$ of 60-300 pJ, and $\tau_c$ of 0.2-1 ns, no single combination of $E_{sat}$ and $\tau_c$ reproduces all curves, based on the simple model of ref. 7. A detailed quantitative analysis of the experimental results would require taking into consideration the amplifier geometry and gain profile, carrier heating, ultrafast dynamics[16], and pulse propagation effects[7,18]. In particular, the fast decrease of gain for TL 25-fs pulses is attributed to smaller values of $E_{sat}$ resulting from the carrier heating dynamics [16].

Figure 3 shows gain curves for input powers <1 mW and also when input pulses are spectrally filtered by the 3-nm bandpass filter at 780 nm placed before the amplifier. In this case, for example, pulses that were chirped to 320 ps and then spectrally filtered, are shortened to 21 ps. The gain for the 21 ps (3 nm) pulse is expected to be significantly smaller than that of the 101 ps (14 nm) pulse due to the shorter pulse duration. On the other hand, the 21 ps pulses experience higher gains since their spectrum is near the peak of the amplifier gain (Fig.1). The combined effects lead to a somewhat reduced gain by 1 - 3 dB with respect to the 101 ps pulses. Fig.3 also shows the curve corresponding to amplification of a spectral window (3 nm at 780 nm) of supercontinuum light generated in a microstructure fiber for stabilizing the carrier-to-envelope offset frequency of the comb. For average input powers limited to 300 μW (within a 3 nm band at 780 nm), we observe gains ranging from 12 to 18 dB. Since the input pulse durations are in between the transform limited femtosecond pulses and the chirped picosecond pulses discussed above, the faster decrease of gain here can also be attributed to smaller $E_{sat}$ due to carrier heating [16]. These gain values demonstrate that TSA can be particularly useful in amplifying a spectral window of supercontinuum light in the near infrared. For example, a commercially available TSA operating at 1.08 μm[15] can be used to boost the power in this region by a factor of 15 dB. This amplification will be beneficial for frequency doubling in a nonlinear crystal used in f-to-2f spectrometers.

To demonstrate phase coherence between the input and amplified frequency comb, an optical heterodyne beat experiment is performed using a Mach-Zender interferometer with the tapered amplifier in one arm and an acousto-optic modulator (AOM) in the other. This enables the optical beat signal to be recorded at a nonzero value with the interferometer path difference set at a zero delay. By mixing the beat signal with another high-quality radio frequency reference, the beat signal is downconverted to an acoustic frequency of a few kilohertz to allow analysis with a high-resolution fast-Fourier-transform analyzer. Figure 4 shows a beat note recorded in a 100-Hz span on the FFT analyzer. The beat signals are monitored as a function of the bias current of the tapered amplifier, with no changes up to 2.4 A. Spectral features up to a few kHz, corresponding to technical noise of electrical and acoustic origins, have also been observed, but do not contribute significantly to reduce the carrier power of the beat. The near spectral-resolution-limited 0.5-Hz linewidth demonstrates that the TSA adds negligible phase noise during the amplification process, and therefore can be used to amplify light from a frequency comb, preserving the phase coherence required for frequency metrology experiments.

In summary, we have demonstrated that TSA can be used for femtosecond frequency combs if the pulses are considerably chirped. TSA preserves the phase coherence of the comb at the $10^{-15}$ level, which is important for optical frequency measurements and optical time transfer in fiber networks. It can be used to amplify a spectral window of supercontinuum light, with advantages for measuring and stabilizing the carrier-to-envelope offset frequency, with very low power in the infrared such as octave-spanning lasers or other low efficiency supercontinuum systems based on microstructure fibers. We expect that semiconductor amplifiers should be equally useful for frequency combs in the telecom region at 1.55 μm, where they compete with fiber amplifiers.

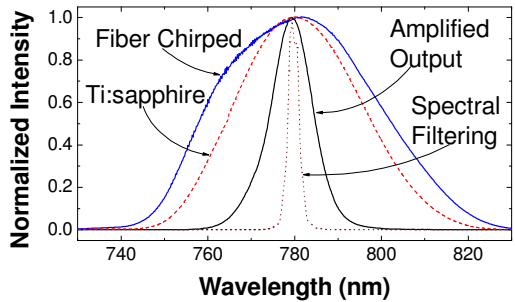

Fig. 1. Red (online) dashed: spectrum of the mode-locked Ti:sapphire laser; Blue (online) solid: laser spectrum after passing through 21 m of optical fiber; black (online) solid: laser spectrum after amplification; purple (online) dotted: transmission profile of 3-nm optical band pass filter at 780 nm.

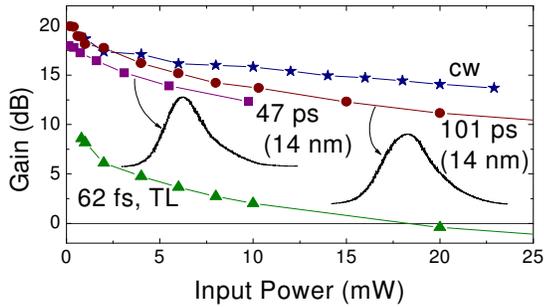

Fig. 2. Gain corresponding to cw (stars) and pulsed injection. Indicated pulse durations are under the TSA BW of 14 nm: 25-fs transform-limited (TL) input pulses widened to 62 fs (▲); pulses chirped to 150 ps (44 nm), then spectrally filtered to 47 fs (■); pulses chirped to 320 ps (44 nm), then spectrally filtered to 101 ps (●). Insets show corresponding output pulse shapes (total x-scale of 200 ps).

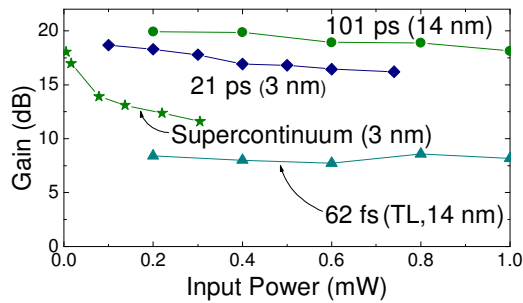

Fig. 3. Small signal gain, with a 3-nm spectral filter inserted before the amplifier. Pulses reaching the amplifier are: TL 25 fs (35 fs), spectrally filtered to 62 fs (14 nm, ▲); pulses chirped to 320 ps (44 nm), then spectrally filtered to 101 ps (14 nm, ●); pulses chirped to 320 ps (44 nm), then spectrally filtered to 21 ps (3 nm, ♦), and from the filtered photonic fiber supercontinuum (stars).

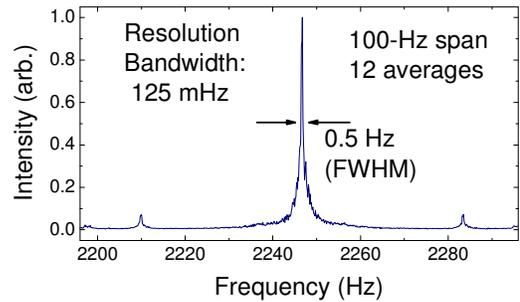

Fig. 4. Beatnote between input and amplified beams, showing sub-Hertz linewidth.

This work is supported by ONR, NASA, and NIST. FCC is a 2005 JILA visiting fellow and he also acknowledges the support of FAPESP and CNPq- Brazil.